\documentclass[sigconf]{acmart}

\usepackage{booktabs} 

\usepackage{makecell}
\usepackage{natbib}

\setcitestyle{square}

\setcopyright{rightsretained}
\copyrightyear{2023}
\acmYear{2023}
\acmConference{SA Posters '23}{December 12-15, 2023}{Sydney, NSW, Australia}
\acmBooktitle{SIGGRAPH Asia 2023 Posters (SA Posters '23), December 12-15,2023}
\acmDOI{10.1145/3610542.3626119}
\acmISBN{979-8-4007-0313-3/23/12}

\begin{document}

\title{Landmark Guided 4D Facial Expression Generation}

\settopmatter{authorsperrow=4}
\author{Xin Lu}

\affiliation{%
  \institution{School of AI, Univ. of CAS}
  \city{Beijing}
  \country{China}
}
\email{luxin22@mails.ucas.ac.cn}

\author{Zhengda Lu}
\affiliation{%
  \institution{School of AI, Univ. of CAS}
  \city{Beijing}
  \country{China}
}
\email{luzhengda@ucas.ac.cn}

\author{Yiqun Wang}
\affiliation{%
 \institution{Chongqing University} 
 \city{Chongqing}
 \country{China}
}
\email{yiqun.wang@cqu.edu.cn}

\author{Jun Xiao}
\authornote{email:xiaojun@ucas.ac.cn(Corresponding authors)}
\affiliation{%
  \institution{School of AI, Univ. of CAS}
  \city{Beijing}
  \country{China}
}
\email{xiaojun@ucas.ac.cn}

\begin{abstract}
In this paper, we proposed a generative model that learns to synthesize the 4D facial expression with the neutral landmark.
Existing works mainly focus on the generation of sequences guided by expression labels, speech, etc, while they are not robust to the change of different identities.
Our LM-4DGAN utilizes neutral landmarks to guide the facial expression generation while adding an identity discriminator and a landmark autoencoder to the basic WGAN for achieving better identity robustness. 
Furthermore, we add a cross-attention mechanism to the existing displacement decoder which is suitable for the given identity.

\end{abstract}

\begin{teaserfigure}
  \includegraphics[width=\textwidth]{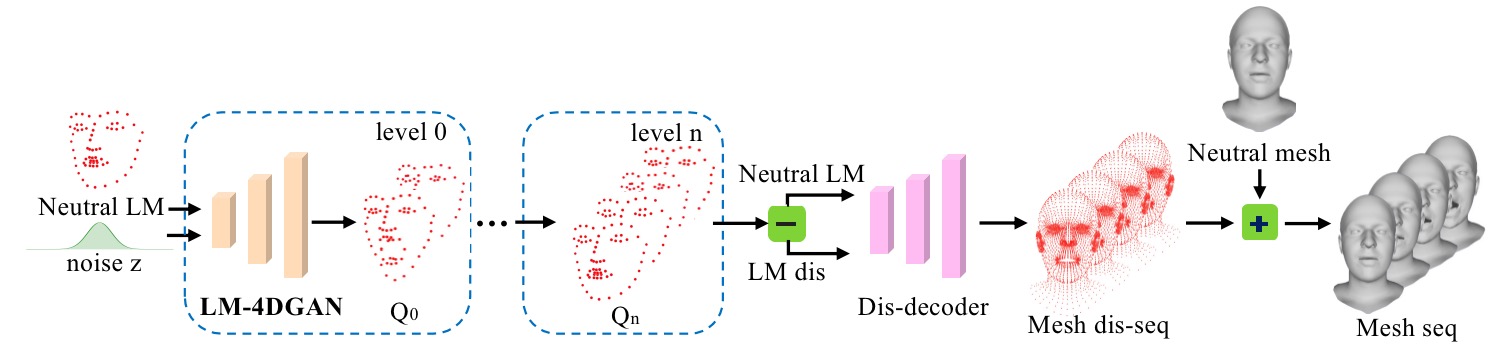}
  \caption{ The overview of our framework. A series of LM-4DGANs synthesize the landmark sequence by using the generated landmark (LM) of the former level (especially, the given neutral LM for the first level) and random noise, then a displacement decoder(dis-decoder) is used to transfer the landmark displacements(LM dis) to the displacement of each mesh vertex(mesh dis-seq). 
  Finally, the 4D facial expression is synthesized by adding these displacements to the neutral 3D mesh. }
  \Description{.}
  \label{fig:pipline}
\end{teaserfigure}

\maketitle

\section{Introduction}
The synthesis of 4D facial expressions is a fundamental and crucial problem in computer vision and graphics, which is widely used in many applications, such as 3D animations, virtual reality, and games. 
This task aims to start from a given neutral landmark which is from neutral mesh by FLAME topology and generate a sequence of realistic 3D face meshes that dynamically evolve across time guided by conditions such as expression labels, speech, etc.
However, it is necessary to use multiple visual sensors to collect dense face mesh sequences as the 4D facial expressions ground truth with local details.
Therefore, there are only a few learning-based methods to study the generation of 4D facial expressions due to the difficulty of data acquisition and the lack of video streaming data.

In prior works, \citep{potamias2020learning} leverages the LSTM to generate the expression latent time-sample and then decode the mesh deformation to add to the neutral expression identity mesh.
Motion3D\citep{otberdout2022sparse} proposes the Motion3DGAN to generate the SRVF coding of the landmark displacements in expression sequence.
Then they decode them into dense mesh vertex displacements by a sparse-to-dense decoder.
These methods establish the relationship between input conditions and the sequences of dense face mesh, while the mesh vertex displacements that they generate are not robust for different face identities.
On the other hand, Motion3DGAN can only generate fixed-length sequences and cannot flexibly synthesize facial expression animations with different lengths.
Therefore, these methods are difficult to be applied to downstream tasks due to these shortcomings.

In this paper, we propose a 4D dynamic facial expression generation network that takes a neutral landmark as guidance to generate vivid facial expressions.
First, we construct a coarse-to-fine architecture based on GANimator\citep{li2022ganimator} to start from random noise and the neutral landmark. Thus, we can generate vivid expressions with variable lengths by a series of LM-4DGANs. 
Our LM-4DGANs are robust to each identity and can generate vivid expressions with variable lengths. 
Meanwhile, we add an identity discriminator and a temporal coherent discriminator to generate realistic landmark displacement sequences.
Then, we conduct a displacement decoder to decode the landmark displacements into dense mesh vertex displacements.
Finally, we can generate more authentic sequences of animated expressions for different identities.

\section{Our approach}

As shown in Fig.\ref{fig:pipline}, our framework consists of a series of LM-4DGANs that generate a landmark expression sequence and a displacement decoder that transfers the landmark displacements into mesh vertex displacements.

In the LM-4DGAN, we construct a coarse-to-fine architecture with neutral landmark as input to guide the synthesis of expression sequences. Then, we can generate vivid expressions with variable lengths while the subsequent level receives the previous output and random noise as input.
Meanwhile, we utilize an autoencoder to encode the landmarks since the sparsity of facial landmarks makes learning their deformations in 3D space difficult.
In the training stage, we add an identity discriminator $D_{iden}$ with the loss $L_{{iden}}$ to discriminate the identity in the generation stage and a temporal coherent discriminator $D_{{coh}}$ with the loss $L_{{coh}}$ to maintain the consistency between consecutive frames.
The loss terms are formulated as:

\begin{equation}
\begin{aligned}
L_{iden}\left(G, D_{{iden}}\right) & =\mathbb{E}_{d \in S(d)}[\log D_{{iden}}(d,LM)] \\
& + \mathbb{E}_{p \in S(p)}\left[\log \left(1-D_{{iden}}\left(G(LM),LM\right)\right]\right.
\end{aligned}
\end{equation}

\begin{equation}
\begin{aligned}
L_{coh}\left(G, D_{{coh}}\right) & =\mathbb{E}_{d \in S(d)}[\log D_{coh}(dif(d))] \\
& + \mathbb{E}_{p \in S(p)}\left[\log \left(1-D_{coh}\left(dif(G(LM))\right)\right]\right.
\end{aligned}
\end{equation}

where the operation $dif$ indicates the deformation between consecutive frames. The parameter $LM$ represents the neutral landmark, which is the input of the generator $G$.

In the decoder part, we add a cross-attention mechanism for the landmark displacements with the neutral landmark to Motion3D’s decoder. Thus, it can be more robust to different identities.

\begin{figure}
  \includegraphics[width=\linewidth]{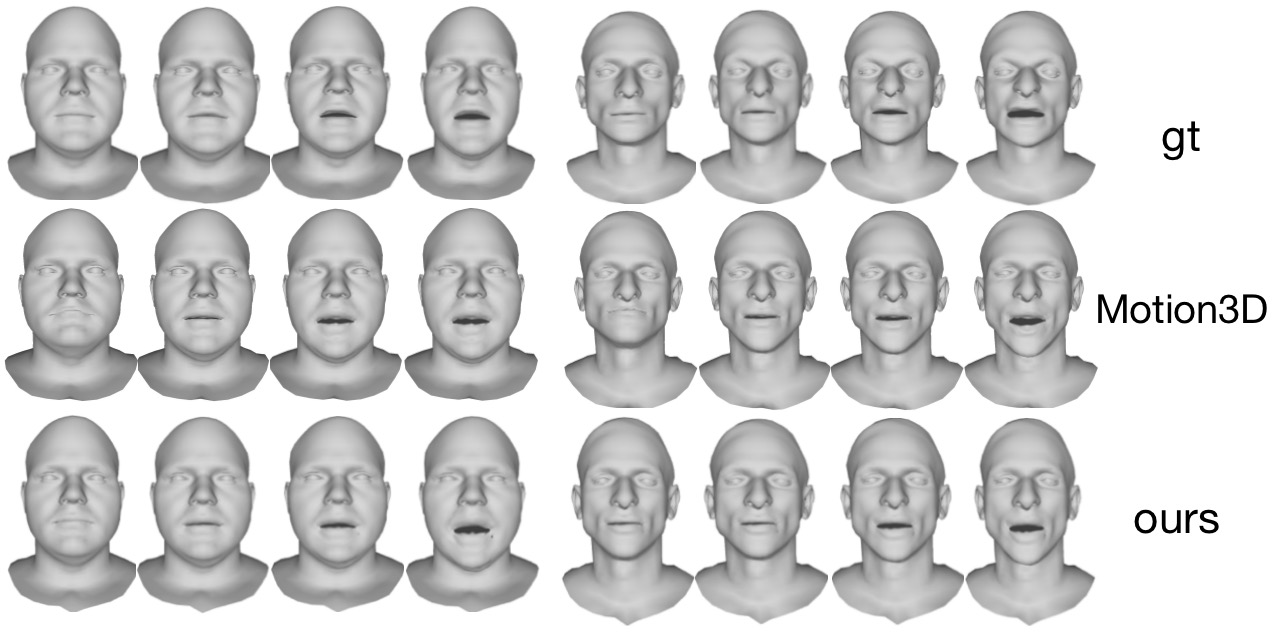}
  \caption{Qualitative results. The first line is the ground truth of the expression mouth-extreme with different identities. The second line is the generating results of Motion3D while the third line is the generating results of our network.}
  \Description{.}
  \label{fig:Quantity}
\end{figure}

\section{EXPERIMENTS AND RESULTS}
We train our LM-4DGANs on CoMA dataset\cite{ranjan2018generating} and the experiments of landmark generation and the mesh displacement decoder are both evaluated with per-vertex reconstruction error(0.1mm). 
Furthermore, the ground truth in our experiments is obtained by interpolating the original data.
Compared with Motion3D, the 4D facial expressions we generate are closer to the ground truth and perform better in detail when given different identities as shown in Fig.\ref{fig:Quantity}.
Then, we conduct a quantitative comparison between our method and Motion3D for the landmark generation. 
Since Motion3DGAN output the SRVF coding of landmark displacement with the different expression labels, we transfer their results into landmark sequences. 
Meanwhile, we generate the same 30-frame expression sequences as the Motion3D for a fair comparison.
The quantitative results are shown in Tab.\ref{tab:comparison}. 
Our method achieves lower per-vertex reconstruction errors in both landmarks and mesh vertexes.
Furthermore, we perform the ablation study for our LM-4DGAN and displacement decoder. 
Our LM-4DGAN greatly improves the accuracy of landmark sequence generation by the autoencoder of landmarks and optimizing the discriminator.
In addition, the cross-attention mechanism in the displacement decoder also effectively improves the result of decoding mesh displacement.

\begin{table}[!htbp]
\centering
    \setlength{\tabcolsep}{1mm}{
     \small
    \begin{tabular}{c|c|c|c|c|c|c}
        \hline

       method \selectfont & Motion3D& ours & w/o $L_{coh}$ & w/o $L_{iden}$ & w/o AE  & w/o atten \\
        \hline
        landmark& 0.750&\pmb{0.562}& 0.583& 0.668& 1.262&{-}\\
        \hline
        mesh& 5.288& \pmb{4.324}& {4.643}&{4.724}&{5.257}& 4.414\\
        \hline

    \end{tabular}}
    \caption{ Quantitative results and ablation studies. These experiment was evaluated by per-vertex reconstruction error(0.1mm).}
    \label{tab:comparison}
\end{table}
\vspace{-0.55cm}

\section{Conclusion}
In this work, we construct a coarse-to-fine architecture to synthesize 4D facial expressions gradually in the time domain which achieve more accurate expression generation results for different identities.
However, with the lack of 4D facial data, we only experiment with our method on the CoMA dataset and will test on other datasets and focus more on temporal indicators in future work.

\begin{acks}
This work is supported by the Fundamental Research Funds for the Central Universities and the Natural Science Foundation of Chongqing (CSTB2022NSCQ-MSX0924).
\end{acks}

\bibliographystyle{ACM-Reference-Format}
\bibliography{ref.bib}

@inproceedings{otberdout2022sparse,
  title={Sparse to dense dynamic 3d facial expression generation},
  author={Otberdout, Naima and Ferrari, Claudio and Daoudi, Mohamed and Berretti, Stefano and Del Bimbo, Alberto},
  booktitle={Proceedings of the IEEE/CVF Conference on Computer Vision and Pattern Recognition},
  pages={20385--20394},
  year={2022}
}

@inproceedings{potamias2020learning,
  title={Learning to generate customized dynamic 3D facial expressions},
  author={Potamias, Rolandos Alexandros and Zheng, Jiali and Ploumpis, Stylianos and Bouritsas, Giorgos and Ververas, Evangelos and Zafeiriou, Stefanos},
  booktitle={Computer Vision--ECCV 2020: 16th European Conference, Glasgow, UK, August 23--28, 2020, Proceedings, Part XXIX 16},
  pages={278--294},
  year={2020},
  organization={Springer}
}

@article{li2022ganimator,
  title={Ganimator: Neural motion synthesis from a single sequence},
  author={Li, Peizhuo and Aberman, Kfir and Zhang, Zihan and Hanocka, Rana and Sorkine-Hornung, Olga},
  journal={ACM Transactions on Graphics (TOG)},
  volume={41},
  pages={1--12},
  year={2022},
  publisher={ACM New York, NY, USA}
}

@inproceedings{ranjan2018generating,
  title={Generating 3D faces using convolutional mesh autoencoders},
  author={Ranjan, Anurag and Bolkart, Timo and Sanyal, Soubhik and Black, Michael J},
  booktitle={Proceedings of the European conference on computer vision (ECCV)},
  pages={704--720},
  year={2018}
}

\end{document}